\documentclass[showpacs,amsmath,amssymb,twocolumn]{revtex4}

\usepackage{graphicx}
\usepackage{dcolumn}
\usepackage{bm}

\begin{document}

\title{\bf\large{Photoproduction of Pentaquark $\Theta^+$ and Chiral Symmetry Restoration\\ in Hot and Dense Medium}}
\author{\normalsize{Xuewen Hao and Pengfei Zhuang }}
\affiliation{Physics Department, Tsinghua University, Beijing
100084, China}

\begin{abstract}
The photoproduction rate of pentaquark $\Theta^+$ is calculated in
a hot and dense medium. At high temperature and density, due to
the restoration of chiral symmetry, photoproduction energy
threshold is increased. Above the thresold the production cross section is strongly enhanced.
\end{abstract}


\maketitle

\section{Introduction}
\label{s1} The pentaquark baryon $\Theta^+ (uudd\bar s)$ candidate
was first observed by the LEPS group at Spring-8 in the $\gamma n
\to K^+ K^- n$ reaction\cite{leps}. However, the existence of the
pentaquark states is remaining an open question as some
experimental groups confirmed the observation \cite{pexp} and some
reported negative results \cite{nexp}. Since Diakonov et al.
predicted the possible existence of the exotic baryon in their
chiral soliton model\cite{diak}, lots of theoretical calculations
for $\Theta^+$ production and related dynamics have been carried
out, see refs\cite{theory}.

At hadron level the process $\gamma n \to \Theta^+ K^-$ has been
analyzed with different couplings\cite{nam}, namely the
pseudoscalar(PS), pseudovector(PV) couplings, and the hybrid
model(HM) which incorporates both PS and PV couplings where PS
dominates the interations at low energy and PV is responsible for
the high energy interactions. Taking into account the finite
hadron size effects, the cross sections for the process are 120 nb
and 15 nb for positive and negative parity $\Theta^+$,
respectively.

More recently the pentaquark production was investigated with
medium effect\cite{chenlw,kim2}. Including the effects of
Quark-Gluon plasma(QGP) formation and hadronic interactions the
authors reported their results of pentaquark production in central
Au+Au collisions at RHIC\cite{chenlw}. They found that the yields
of $\Theta^+$ is determined by the initial formation from QGP
phase and the hadronic rescatterings do not affect the total cross
sections remarkably.

It is well known that the spontaneous breaking of chiral symmetry
controls the vacuum properties of hadrons. The restoration of the
symmetry at finite temperature and density is one of the most
important predictions of the lattice gauge theory(LGT)
calculation. It is found that\cite{our1} the effect of the
symmetry restoration on the pentaquark production is significant
and sensitive to the parity of the pentaquark. In the hot and
dense medium, the shift in mass $\Delta M_\Theta$, and the shift
in width $\Delta \Gamma_\Theta$ are found to be much larger for
positive parity $\Theta^+$ than that for negative parity. While
the magnitude of the mass shift is much smaller than the mass
$M_\Theta =$ 1540 MeV in vacuum, the shift in width is several
times larger than the vacuum width. On the other hand, in the
absence of the chiral restoration, the shifts in mass and width
can be neglected.

In this paper, we present the results of photoproduction rate of
$\Theta^+$ in medium. The effects of the chiral symmetry
restoration at high temperature and density on the production will
be analyzed. The formula for $\Theta^+$ production in medium,
including the PS, PV and HM couplings, will be presented in
Section \ref{s2}. Results and discussions are in Section \ref{s3}
and the summaries will be given in the last Section.

\section{Formulas}
\label{s2} We take the effective pseudovector and pseudoscalar
$N\Theta^+ K$ couplings\cite{nam,kim2},
\begin{eqnarray}
\label{lag}
{\cal L}_{PV} &=&-\frac{g_A^*}{2 f_\pi} \bar\Theta
\gamma_\mu\gamma_5\left(
\partial^\mu K^+ n -\partial^\mu K^0 p\right)\ , \nonumber\\
{\cal L}_{PS} &=&ig \bar\Theta \gamma_5 \left(K^+ n - K^0 p\right)
\ ,
\end{eqnarray}
where, the positive parity of $\Theta^+$ is assumed. Since the
parity of $\Theta^+$ has not yet been determined experimentally,
we will consider in the following the interactions with negative
parity of $\Theta^+$ by removing the factor $ i\gamma_5$ from the
above vertexes. The pseudovector and pseudoscalar coupling
constants $g_A^*$ and $g$ can be fixed by reproducing the mass
$M_\Theta =$ 1540 MeV and decay width $\Gamma_\Theta =$ 15 MeV in
the vacuum. Through the calculation at tree level one has $g_A^*=$
0.28 and $g=$ 3.8 for positive parity and $g_A^*=$ 0.16 and $g=$
0.53 for negative parity\cite{nam}.

The in-medium static properties, especially the particle masses,
are different from that in vacuum. It is well known that the
in-medium behavior of nucleons can be well described by the chiral
symmetry restoration at finite temperature and density. While the
quantitative result depends on the models used, the qualitative
temperature and density dependence of the nucleon mass $M_N$ is
not sensitive to the details of different chiral
models\cite{brown}. A simple model to describe chiral symmetry
breaking in vacuum and symmetry restoration in medium is the
Nambu--Jona-Lasinio (NJL) model\cite{njl}. Within this model, one
can obtain the hadronic mass spectrum and the static properties of
mesons remarkably well. In particular, one can recover the
Goldstone mode, and some important low-energy properties of
current algebra such as the Goldberger-Treiman and
Gell-Mann-Oakes-Renner relations\cite{vogl}. In mean field
approximation of NJL, the effective nucleon mass is determined
through the quark gap equation\cite{vogl,our2},
\begin{eqnarray}
\label{njl}
&& M_N = 3 m_q\ ,\nonumber\\
&& 1-4N_cN_fG\int{d^3{\bf p}\over (2\pi)^3}{1\over E_p}
\Big[1-f_f\left(E_p + {\mu\over 3}\right)\nonumber\\
&&-f_f\left(E_p-{\mu\over 3}\right)\Big] = {m_0\over m_q}\ ,
\end{eqnarray}
where $\mu$ is the baryon chemical potential, $T$ is the
temperature, $N_c$ and $N_f$ are the color and flavor degrees of
freedom, $m_q$ and $E_p = \sqrt{p^2 + m_q^2}$ are the effective
quark mass and energy, and $f_f$ is the Fermi-Dirac distribution
function $f_f(x)=1/\left(e^{x/T}+1\right)$. The three parameters
in the model, the current quark mass $m_0$ which breaks explicitly
the chiral symmetry in vacuum, the coupling constant $G$, and the
momentum cutoff $\Lambda$, can be fixed by fitting the chiral
condensate, pion mass and pion decay constant in
vacuum\cite{vogl,our2}. The temperature and density dependence of
the nucleon mass calculated through (\ref{njl}) is very close to
that obtained from the other often used models describing nuclear
matter\cite{brown}.

Since the effects of the attractive scalar and the repulsive
vector potentials cancel each other, the mass of $K^+$ only
increase slightly at finite temperature and density. Although the
$K^-$ mass drops much larger in the medium, it is negligible
compared to the changes of in-medium nucleon mass. For these
reasons, we take the kaon mass as a constant through out our
calculations\cite{kim2,our1}.

The temperature and density effect on the pentaquark $\Theta^+$
mass and width can be determined through calculating the lowest
order $\Theta^+$ self-energy in the medium\cite{our1}.

Now we come to the in-medium calculation of the cross section for
the process $\gamma n \to \Theta^+ K^-$ shown diagrammatically in
Fig.\ref{fig1}. To have a comparison with the calculations in
vacuum\cite{nam,oh}, we take only the diagrams at tree level. For
the channel PS, we consider the Born terms Fig.\ref{fig1}b, c and
d, while for the channel PV, we should include the contact term
Fig.\ref{fig1}e because of the replacement
$\partial_\mu\rightarrow
\partial_\mu+ieA_\mu$ in the lagrangian (\ref{lag}).

The electromagnetical vertexes in Fig.\ref{fig1} are defined
through the Lagrangian density\cite{nam}
\begin{widetext}
\begin{center}
\begin{figure}[!htb]
\begin{center}
\includegraphics[width=12cm]{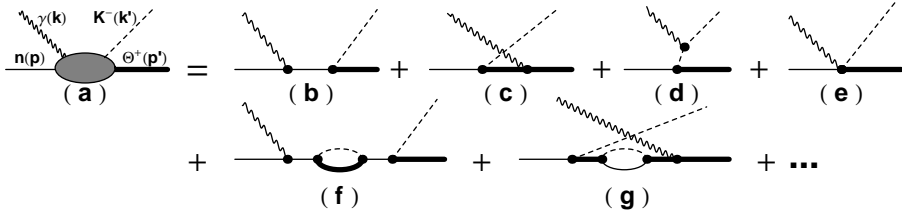}
\caption{The diagrams for the process $\gamma n \to \Theta^+ K^-$.
\label{fig1}}
\end{center}
\end{figure}
\end{center}
\end{widetext}

\begin{eqnarray}
{\cal L}_{\gamma KK}&=&ie\left[\left(\partial^{\mu}K^-\right)K^+-\left(\partial^{\mu}K^+\right)K^-\right]A_{\mu}\ ,\nonumber \\
{\cal L}_{\gamma nn}&=&-i\frac{e\kappa_n}{2M_n}\bar{n}\sigma_{\mu \nu}k^{\nu}n A^{\mu}\ ,\nonumber \\
{\cal L}_{\gamma \Theta
\Theta}&=&-e\bar{\Theta}\left(\gamma_{\mu}+i\frac{\kappa_{\Theta}}
{2M_{\Theta}}\sigma_{\mu \nu}k^{\nu}\right)\Theta A^{\mu}\ ,
\end{eqnarray}
where $\kappa_n$ and $\kappa_\Theta$ are the neutron and
pentaquark magnetic moments in unit of nuclear magneton, we take
$\kappa_n = -1.91$ and $\kappa_\Theta=0$\cite{nam} through out the
calculation.

In the initial nucleon rest frame, the cross section at tree level
in the medium can be written as
\begin{eqnarray}
\label{cross}
\sigma(k) &=& \int_0^{\theta_{\mathrm {max}}}{d
\theta \sin \theta\over 32\pi} \frac{|{\mathbf k'}|^2}{kM_N}
\frac{1}{\big |
(k+M_N) |{\mathbf k'}|-E_{K^-} k \cos \theta  \big|}\nonumber\\
&\times& \sum |i {\cal M}|^2 \left(1 -
f_f(E_{\Theta}-\mu)\right)\left( 1 + f_b(E_{K^-})\right)\ ,
\end{eqnarray}
where ${\cal M} = {\cal M}_{\gamma n \to \Theta^+ K^-}$ is the
amplitude of the process, $\theta$ is the angle between the
initial photon and final kaon momenta, $\theta_{max}$ is the
maximum angle restricted by energy and momentum conservations, and
$E_{K^-} = \sqrt{{\bf k'}^2 + M_{K^-}^2}$ and $E_{\Theta} =
\sqrt{{\bf p'}^2 + M_{\Theta}^2}$ are $K^-$ and $\Theta^+$
energies. The summation symbol in (\ref{cross}) means the average
over the initial states (the photon polarization and neutron spin)
and the summation over the final states (the pentaquark spin). The
temperature and density effect in the cross section is hidden in
the nucleon and pentaquark masses which are governed by the chiral
symmetry restoration and shown explicitly in the final state
statistics factor $(1-f_f)(1+f_b)$ with the Bose-Einstein
distribution function $f_b(x) = 1/\left(e^{x/T}-1\right)$.

For the hybrid model (HM), which combines the PS and PV channels,
can better reproduce the experimental result than the individual
PS or PV channel in vacuum, we use the following
amplitude\cite{nam},
\begin{eqnarray}
{\cal M}_{HM} & = & {\cal M}_{PS} + \Delta {\cal M} \cdot F_{HM}\ , \nonumber \\
\Delta {\cal M} & = & {\cal M}_{PV} - {\cal M}_{PS}\ ,
\end{eqnarray}
where $F_{HM} = \Lambda_{HM}^2 / \left( \Lambda_{HM}^2 + {\bf
k'}^2 \right)$ is the mixing parameter depending on the kaon
momentum ${\bf k'}$ and the cutoff $\Lambda_{HM}=$ 450 MeV
\cite{nam,drechsel}.

\section{Numerical Results}
\label{s3} The results of the in-medium pentaquark production
cross section, as a function of photon energy $E_\gamma =k$, are
shown in Fig.\ref{fig2}. Both positve and negative parities of the
pseudoscalar(PS), pseudovector(PV) and hybrid couplings(HM) are
shown in left- and right-plots, respectively. In each case, we
consider three different properties of the medium characterized by
temperature T and baryon chemical potential $\mu$. The values of
$(T,\mu)=(0.2, 0.05)$ GeV (dotted lines), $(0.15,0.3)$ GeV (dashed
lines) and $(0.1, 0.8)$ GeV (dot-dashed lines) are approximately
corresponding to the fireballs created in heavy ion collisions at
RHIC, SPS, and AGS, respectively. The results corresponding to the
vacuum are also shown as solid-lines in the plots\cite{nam,oh}. In
all calculations only baryon chemical potential is considered and
the strangeness chemical potential is ignored. Therefore the
pentaquark chemical potential is the same as the nucleon chemical
potential. Since the photon energy is in the range of 10 GeV in
heavy ion collisions\cite{photon}, we limited our calculation at
12 GeV.

We first discuss the production threshold:
\begin{equation}
\label{thre} k_{th}={\left( M_{\Theta}+M_K \right)^2-M_N^2 \over
2M_N}\ .
\end{equation}

\begin{widetext}
\begin{center}
\begin{figure}[!htb]
\begin{center}
\includegraphics[width=12cm]{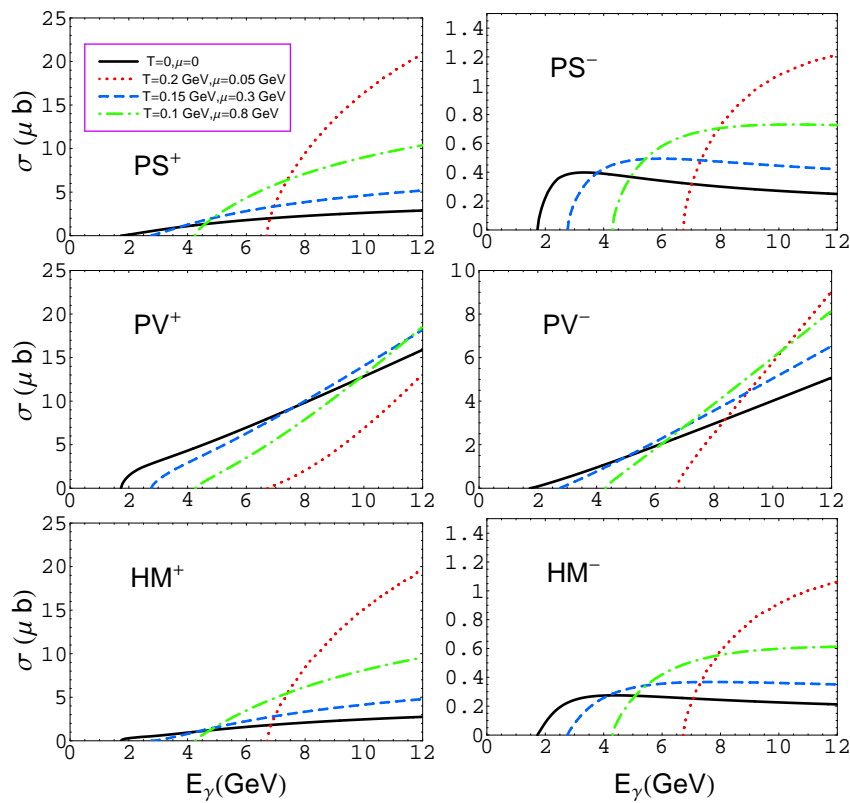}
\caption{The pentaquark production cross section for the process
$\gamma n \to \Theta^+ K^-$ as a function of photon energy for six
kinds of interactions, pseudoscalar, pseudovector and hybrid
couplings with positive and negative parity $\Theta^+$ (PS$^+$,
PS$^-$, PV$^+$, PV$^-$, HM$^+$, and HM$^-$). The solid, dotted,
dashed and dot-dashed lines correspond, respectively, to the
mediums with $(T, \mu)=(0,0)$, $(0.2, 0.05)$ GeV, $(0.15, 0.3)$
GeV, and $(0.1, 0.8)$ GeV. \label{fig2}}
\end{center}
\end{figure}
\end{center}
\end{widetext}

\noindent It is found that the dependence on the temperature and
potential of the nucleon mass\cite{vogl,our2} is much stronger
than that of the pentaquark mass\cite{our1}. At finite temperature
and potential, the decreased nucleon mass leads to the increase of
the $k_{th}$. From the equation (\ref{njl}), we have
$M_N(T,\mu)/M_N(0,0)=0.32, 0.71, 0.49$ corresponding to the
mediums with $(T, \mu)=(0.2, 0.05)$ GeV, $(0.15, 0.3)$ GeV, and
$(0.1, 0.8)$ GeV, the maximum shift is for $(T, \mu)=(0.2, 0.05)$
GeV and the minimum shift is for $(T, \mu)=(0.15, 0.3)$ GeV. Since
the nucleon mass is independent of either the coupling or the
parity, the shifts are the same for all couplings and parities.

As one can see in Fig.\ref{fig2}, when above the threshold, the
production cross section in the medium is enhanced compared to the
vacuum production. In the cases of PS and HM, the enhancement is
universal at any temperature and density. Similar to the threshold
shift, the enhancement strongly depends on the nucleon mass drop,
i.e. the degree of chiral symmetry restoration.

As we mentioned above, the temperature and density are introduced
into our calculation through two ways. The first one is through
the effective nucleon mass which is dominated by the degree of
chiral symmetry restoration. The second one, independent of the
chiral properties, is the statistical factor $(1-f_f)(1+f_b)$ and
the summation of the loop frequency when the in-medium pentaquark
mass is evaluated. In order to make sure that the chiral symmetry
restoration is the driving force for the threshold shift and the
enhancement in the production cross section, we now repeat the
cross section calculations, for PS$^+$ coupling, without the
chiral transition and forcing the nucleon mass at its vacuum
value. The result is shown in Fig.\ref{fig3}. For the other
channels the results are similar. Qualitatively different from the
case with chiral symmetry restoration shown in Fig.\ref{fig2}
where the threshold shifts to the right significantly, it shifts
now to the left slightly due to the small decrease of the
in-medium pentaquark mass resulted from the loop frequency
summation. From the ratio of the cross section without chiral
restoration to the one in the vacuum, shown in the right panel of
Fig.\ref{fig3}, the in-medium effect on the cross section can be
neglected, especially in the high energy region, if the chiral
phase transition is turned off.

\begin{widetext}
\begin{center}
\begin{figure}[!htb]
\begin{center}
\includegraphics[width=14cm]{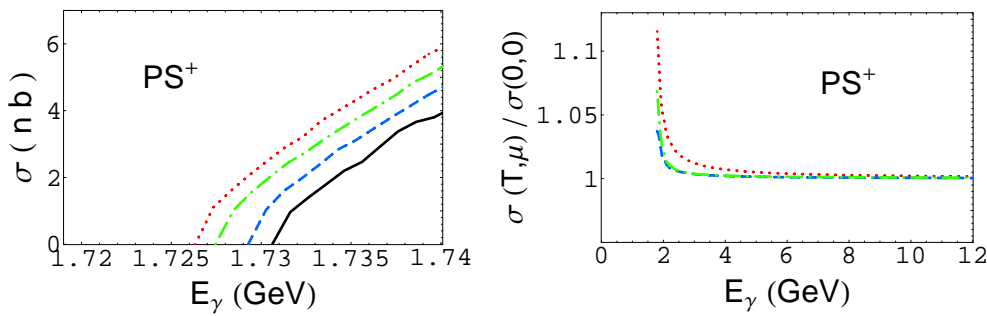}
\caption{The cross section without considering the chiral symmetry
restoration for the channel PS$^+$. The left panel is the cross
section around the threshold, and the right panel is the ratio of
the cross section in the medium to the one in the vacuum. The
solid, dotted, dashed and dot-dashed lines correspond,
respectively, to the mediums with $(T, \mu)=(0,0)$, $(0.2, 0.05)$
GeV, $(0.15, 0.3)$ GeV, and $(0.1, 0.8)$ GeV. \label{fig3}}
\end{center}
\end{figure}
\end{center}
\end{widetext}

We also analyzed the dependence of the production cross section on
magnetic moment $\kappa_\Theta$ and width $\Gamma_\Theta$. The
hadron masses and the production threshold $k_{th}$ are
independent of $\kappa_\Theta$, but the amplitude ${\cal M}$ is
related to $\kappa_\Theta$ due to the vertex $\gamma\Theta\Theta$.
The estimated range of the magnetic moment is
$0<\kappa_{\Theta}<0.7$\cite{theory,nam,oh}. A 50\% increase in
the production cross section is found in our calculation changing
$\kappa_\Theta$ from zero to 0.7 at $E_\gamma =$ 12 GeV with
PV$^-$ coupling. In general, we find that the production cross
section is increased with $\kappa_\Theta$ for negative parity and
decreased with $\kappa_\Theta$ for positive parity $\Theta^+$. In
all above calculations, a fixed width $\Gamma_\Theta =$ 15 MeV has
been used\cite{nam}. When we change $\Gamma_\Theta$, the nucleon
mass does not change and the pentaquark mass shift in the medium
is always small compared with its vacuum value $M_\Theta =$ 1540
MeV. When we neglect the in-medium mass shift we reach the
following relationship between production cross sections:
\begin{equation}
\label{width} {\sigma_{\Gamma_\Theta^1}\over
\sigma_{\Gamma_\Theta^2}} \sim {\Gamma_\Theta^1\over
\Gamma_\Theta^2}\ .
\end{equation}

\section{Conclusions}
\label{s4} We reported our investigation on the production of
pentaquark $\Theta^+$ through the process $\gamma n \to \Theta^+
K^-$. Medium effects via PS, PV, HM couplings were included in the
study. The NJL model was used to describe the mean field with
chiral degrees of freedom. At high temperature and density, we
found that the restoration of chiral symmetry has two significant
effects on the pentaquark production rates: (1) The threshold
shifts towards higher energy; (2) Beyond the threshold, the total
cross section is strongly enhanced. Without the chiral phase
transition, the medium effect on the pentaquark production is
found to be negligible.

\vspace{0.3in}

\noindent {\bf \underline{Acknowledgments:}} We thank Xuguang
Huang for helpful discussions and Nu Xu for careful reading of the
manuscript. The work is supported in part by the grants
NSFC10435080,10428510 and SRFDP20040003103.

\end{document}